\documentclass[%
 reprint,
 superscriptaddress,
 amsmath,amssymb,
 aps,
 floatfix,
]{revtex4-2}

\usepackage{comment}
\usepackage{eqnarray}
\usepackage{physics}
\usepackage{dsfont}
\usepackage{tensor}
\usepackage{amsthm}
\usepackage{amsmath}
\usepackage{bbold}
\usepackage{fancyhdr}
\usepackage{subfigure}
\usepackage{epsfig}
\usepackage{bbm}
\usepackage{subfigure}
\usepackage{bm}
\usepackage{tikz}
\usepackage{float}
\usepackage{xcolor}
\usepackage{amsmath}
\usepackage{graphicx}
\usepackage{dcolumn}
\usepackage{bm,braket}
\usepackage{comment}

\usepackage{soul}

\renewcommand{\selectlanguage}[1]{}
\usepackage{hyperref}

\begin{document}

\preprint{}

\title{Transport of molecules via polymerization in chemical gradients}

\author{Shashank Ravichandir}
\thanks{These two authors contributed equally}
\affiliation{Institut Theory der Polymere, Leibniz-Institut f\"ur Polymerforschung Dresden,  01069 Dresden, Germany}
\author{Bhavesh Valecha}
\thanks{These two authors contributed equally}
\affiliation{Institut f\"ur Physik, Universit\"at Augsburg, 86159 Augsburg, Germany}
\author{Pietro Luigi Muzzeddu}
\affiliation{Department of Biochemistry, University of Geneva, 1211 Geneva, Switzerland}%
\author{Jens-Uwe Sommer}
\email[Corresponding author: ]{jens-uwe.sommer@tu-dresden.de}
\affiliation{Institut Theory der Polymere, Leibniz-Institut f\"ur Polymerforschung Dresden,  01069 Dresden, Germany}
\affiliation{Institut f\"ur Theoretische Physik, Technische Universit\"at Dresden, 01069 Dresden, Germany}
\author{Abhinav Sharma}
\email[Corresponding author: ]{abhinav.sharma@uni-a.de}
\affiliation{Institut Theory der Polymere, Leibniz-Institut f\"ur Polymerforschung Dresden,  01069 Dresden, Germany}
\affiliation{Institut f\"ur Physik, Universit\"at Augsburg, 86159 Augsburg, Germany}

\date{\today}

\begin{abstract}
The transport of molecules for chemical reactions is critically important in various cellular biological processes. Despite thermal diffusion being prevalent in many biochemical processes, it is unreliable for any sort of directed transport or preferential accumulation of molecules. In this paper we propose a strategy for directed motion in which the molecules are transported by active carriers via polymerization. This transport is facilitated by chemical/activity gradients which generate an effective drift of the polymers. By marginalizing out the active degrees of freedom of the system, we obtain an effective Fokker-Planck equation for the Rouse modes of such active-passive hybrid polymers. In particular, we solve for the steady state distribution of the center of mass and its mean first passage time to reach an intended destination. We focus on how the arrangement of active units within the polymer affect its steady-state and dynamic behaviour and how they can be optimized to achieve high accumulation or rapid motility.
\end{abstract}

\maketitle

Chemical gradients are ubiquitous in biological systems across length scales. These gradients help in carrying out chemical reactions, mechanical work, and biological interactions \cite{muller1998gradients}, which include growth and migration of cells, healing of wounds, cancer metastasis \cite{keenan2008biomolecular}, and in the positioning of nuclei within cells \cite{almonacid2015active}. Polymers also play a prominent role in microbiological processes. These include DNA transcription and replication where DNA/RNA polymerases move along the DNA \cite{guthold1999direct,alberts2000molecular}, the dynamics of chromosomal loci \cite{javer2013short, weber2012nonthermal} and chromatin\cite{zidovska2013micron}, arrangement of eukaryotic genome \cite{di2018anomalous}, among many others. It has also been shown that polymerization itself is crucial for many cellular processes such as the formation of cell organelles via phase separation \cite{banani2017biomolecular, sommer2022polymer}.

This work seeks to highlight another possible function of polymerization - transport of molecules to their intended locations by means of active carriers \cite{ramaswamy2010mechanics, ramaswamy2017active, de2015introduction, fodor2016far, gompper20202020, julicher2018hydrodynamic, marchetti2013hydrodynamics} in chemical/activity gradients. Such active-passive hybrid polymers are idealized models of biological filaments and microtubules which are active at certain locations due to molecular motors like myosin and kinesin \cite{brangwynne2008cytoplasmic,lu2016microtubule, ravichandran2017enhanced,weber2015random}. It is important to recognize that the term active polymers is also used for polymers in non-equilibrium surroundings like bacterial baths and there have been various studies that look into the structural and dynamical properties of both interpretations of active polymers \cite{winkler2017active, winkler2020physics, anand2018structure, isele2015self, philipps2022tangentially, locatelli2021activity,kaiser2014unusual, zhang2023configurational, anderson2022polymer, harder2014activity, mousavi2021active, shin2015facilitation, foglino2019non, bianco2018globulelike}. However, it is only recently that self-localization of these polymers in response to chemical gradients has received some attention \cite{vuijk2021chemotaxis,muzzeddu2024migration}. In this work we 
consider 
polymer molecules that are inherently active and self-propelled due to a fuel/activity field.

\begin{figure}[h]
		\centering
		\includegraphics[width=0.45\textwidth]{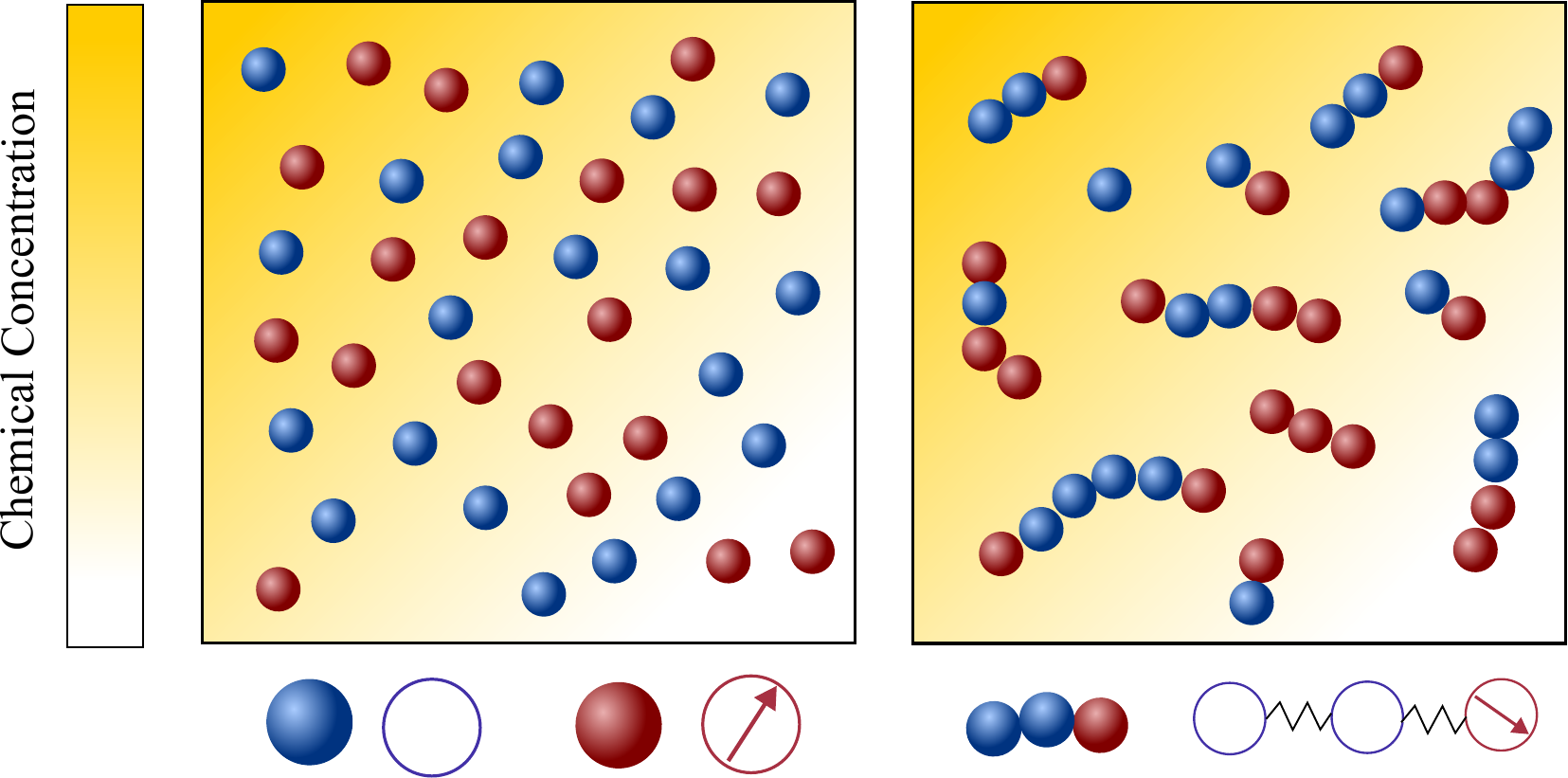}
		\caption{\label{fig:mixture}(Left) Schematic diagram of a mixture of active (red spheres) and passive (blue spheres) particles in an inohmogeneous environment. These spheres can either be representative of colloidal particles or coarse grained molecules. (Right) After undergoing polymerization these particles form composite polymers which have a mixture of active and passive monomers. The polymers are considered to be Rouse chains, with particles connected to each other via simple springs. This is a snapshot of a dynamic system where the positions of polymers do not reflect their steady state localization with respect to the chemical concentration and are only used for representative purposes. }
	\end{figure}

Consider an environment, in which there are gradients in the fuel concentration, filled with a mixture of active (the carriers) and passive (the molecules to be delivered to their intended destinations) monomers that can polymerize linearly into active-passive hybrid polymers as shown in Fig. \ref{fig:mixture}. The passive and active monomeric units are modelled as simple Brownian particles and active Brownian particles (ABPs), respectively. The choice of ABPs is motivated by the fact that it is a simple model that is also used to describe self-propelled colloidal molecules \cite{lowen2018active, elgeti2015physics, bechinger2016active} that can be synthesized in labs \cite{dreyfus2005microscopic, najafi2004simple, jiang2010active, theurkauff2012dynamic, valadares2010catalytic, howse2007self}. It has been analytically demonstrated that individual ABPs, whose swim speeds are proportional to the local fuel concentration \cite{howse2007self}, accumulate in regions of low fuel concentration or activity\cite{schnitzer1993theory,sharma2017brownian}. This is in contrast to some living systems, e.g. the bacterium E. Coli which moves up the activity gradient by altering its tumble rate \cite{berg2004coli}, thus leading to chemotaxis \cite{schnitzer1990strategies,berg1993random}. Though there have been some studies on the transient behavior of ABPs leading to the phenomenon of "pseudochemotaxis" \cite{lapidus1980pseudochemotaxis,peng2015self,ghosh2015pseudochemotactic,vuijk2018pseudochemotaxis,merlitz2020pseudo}, this did not lead to any sort of preferential accumulation in regions of high activity.  However, recent works\cite{vuijk2021chemotaxis, vuijk2022active,muzzeddu2022active,muzzeddu2023taxis,valecha2024chirality} have shown that connected structures with single or multiple ABPs can show chemotactic behaviour, the principles of which we'll be building upon. 

In the present work, we address the following questions: i. does the number and location of active monomers in a polymer affect its preferential accumulation? ii. Are the polymer chains that localize most effectively also the fastest in getting to the target region? Investigating these will help us understand the static and dynamic accumulative behavior of active-passive hybrid polymers. We also show that the results of this work are unaffected by the model of activity of the particles, and hence, could give insights into the transport of molecules for various bio-chemical processes.
	
We model the active-passive hybrid polymers as Rouse chains, in which the interacting monomers are connected to each other via harmonic springs of stiffness $\zeta$. 
For a polymer of chain length $N$, the activities are assigned by the binary variables $\lbrace \alpha_i \rbrace$, with $i = \lbrace 0,1\dots N-1\rbrace$. Specifically, $\alpha_i = 1$ if the $i$-th monomer is active and $\alpha_i = 0$ otherwise.
The overdamped Langevin equations describing the stochastic dynamics of the positions $\lbrace \bm{X}_i (t)\rbrace$ and orientation unit vectors $\lbrace \bm{p}_i (t)\rbrace$ of individual monomers are:
    \begin{equation}
        \label{eqn:langevin}
        \begin{split}
            \bm{\dot{X}}_i(t) &= -\mu \nabla_{\bm{X}_i}\mathcal{H} + \mu \alpha_i f_s(\bm{X}_i)\bm{p}_i + \bm{\xi}_i(t),\\
		\bm{\dot{p}}_i(t) &= \bm{p}_i \times \bm{\eta}_i(t),
        \end{split}
	\end{equation} 
 where $\mu$ is the mobility of the particle, $f_s$ is the swim force due to the spatially varying fuel concentration or activity field, $\lbrace \bm{\xi}_i (t)\rbrace$ and $\lbrace \bm{\eta}_i (t)\rbrace$ are zero-mean white Gaussian noises with correlations $\langle\bm{\xi}_i(t) \otimes \bm{\xi}_j(s)\rangle = 2D\bm{I}\delta_{ij}\delta (t-s)$ and $\langle\bm{\eta}_i(t) \otimes \bm{\eta}_j(s)\rangle = 2D_r\bm{I}\delta_{ij}\delta (t-s)$. Here $\otimes$ represents the outer product, $D$ and $D_r$ are the translational and rotational diffusive coefficients, respectively, and $\bm{I}$ is the $d\times d$ identity matrix, $d$ being the number of dimensions. The Hamiltonian $\mathcal{H}$ modeling the spring interactions between particles is
 \begin{equation}
		\mathcal{H} = \frac{\zeta}{2}\sum_{ij}M_{ij}\bm{X}_i\cdot\bm{X}_j,
\end{equation}
where $M_{ij}$ is the connectivity matrix \cite{sommer1995statistics} for the polymer chain. Even though the relations we derive are applicable for a general connectivity matrix, we will be restricting our study to linear polymer chains, i.e $M_{ij}$ is a tridiagonal matrix.  We write down the Fokker-Planck equation \cite{gardiner1985handbook, risken1996fokker} in terms of the Rouse modes $\lbrace \bm{\chi}_i \rbrace$ \cite{doi1988theory} obtained via the transformation $\bm{\chi}_i = \sum_{j} \varphi_{ij} \bm{X}_j$, where $\varphi_{ij}$ is the diagonalizing matrix of $M_{ij}$ such that $\sum_{jk} \varphi_{ij}M_{jk}\varphi^{-1}_{kl} = \frac{\gamma_i}{\gamma}\delta_{il}$. $\gamma_i$'s are the relaxation rates of the individual Rouse modes and they are normalized by the relaxation rate due to the harmonic interactions $\gamma = \mu\zeta$. The coarse-grained Fokker-Planck equation  for the probability density $\rho (\bm{X}_{\rm{COM}}, t)$ in terms of the polymer's center of mass, $\bm{X}_{\rm{COM}} = \bm{\chi}_0/\sqrt{N}$, is then obtained by integrating out the orientation vectors $\lbrace \bm{p}_i (t)\rbrace$ and the other Rouse modes $\lbrace \bm{\chi}_i (t); i\neq 0 \rbrace$ under a small gradient 
 approximation as-
    \begin{equation}
    \label{eqn:conservativeeqn}
        \frac{\partial \rho}{\partial t} = -\nabla\cdot (\rho \bm{\mathcal{V}} - \nabla (\mathcal{D}\rho)),
    \end{equation}
    where the gradients are with respect to the center of mass coordinates. The complete Fokker-Planck equation and the coarse-graining procedure can be found in the supplementary material. The effective drift $\bm{\mathcal{V}}$ and diffusive coefficient $ \mathcal{D}$ are functions of $\bm{X}_{\rm{COM}}$ and are given by 
    \begin{equation}
    \label{eqn:coefficients}
    \begin{split}
        \bm{\mathcal{V}}(\bm{X}_{\rm{COM}}) &= \frac{\tau}{dN}\left(\frac{S_1 + S_2}{2}\right)\nabla\left(v^2\left(\bm{X}_{\rm{COM}} \right)\right)\\
        \mathcal{D}(\bm{X}_{\rm{COM}}) &= \frac{1}{N}\left(D + \frac{\tau}{d}S_2v^2\left(\bm{X}_{\rm{COM}}\right)\right)
    \end{split}
    \end{equation}
with $\tau = 1/[(d-1)D_r]$, and $v(\bm{x})$ being the swim speed of ABP's ($v = \mu f_s$). $S_1$ and $S_2$ are given by -
    \begin{equation}
    \label{eqn:S}
        S_1 = \sum_{l=1,j=0}^{N-1}\frac{1}{1+\tau\gamma_l} \varphi_{lj}^2 \alpha_j^2, \;\;\;\;\;
	S_2 = \frac{1}{N}\sum_{j=0}^{N-1} \alpha_j.
    \end{equation}
    $S_2$ is the fraction of monomers that are active in the polymers while $S_1$ is a summation that involve the eigenvector matrix $\phi_{ij}$ and the re-scaled eigenvalues $\lbrace \gamma_i \rbrace$. 

	\begin{figure}[h]
		\centering
		\includegraphics[width=0.45\textwidth]{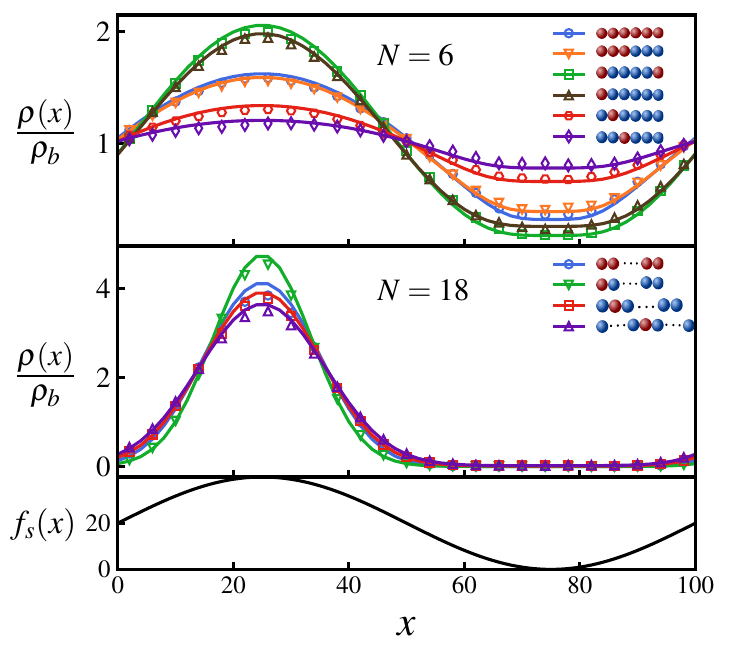}
		\caption{\label{fig:steadystateprofiles}  The steady state density profiles for various configurations of a polymer with chain length $N=6$ and $N=18$ for $k = 8$, $D_r = 5$, $\mu = 1$, and $D=1$ in a box of length $L=100$ that has a sinusoidally varying  activity profile in the x-direction: $f_s(x) = 20\left(1 + sin\left(\frac{2\pi x}{L} \right) \right)$ as depicted. The densities are normalized by  $\rho_b = 1/L$. The solid lines represent analytical predictions and the symbols represent Langevin dynamics simulation results.}
	\end{figure}
	
	One can see in Eq. \eqref{eqn:coefficients} that the diffusivity of the polymer is enhanced due to activity compared to its all-passive counterpart (in which case the effective diffusivity would have been $D/N$) and the enhancement is proportional to the fraction $S_2$ of active monomers in the polymer. The effective drift $\mathcal{V}$ is generated by the gradient of the swim force squared, and  is otherwise absent in passive or even constant activity systems. The steady state density of the center of mass of the polymer chain can now be calculated using the zero flux condition in \eqref{eqn:conservativeeqn} and the fact that $\bm{\mathcal{V}} =(1  -\frac{\epsilon}{2})\nabla \mathcal{D}$, where 
    \begin{equation}
    \label{eqn:epsilon}
		\epsilon = \frac{S_2 - S_1}{S_2}.
	\end{equation}
    In particular, we obtain-
	\begin{equation}
		\label{eqn:steadystate}
		\rho(\bm{X_{\rm{COM}}}) \propto \left[1 + \frac{\tau S_2v^2\left(\bm{X_{\rm{COM}}}\right)}{dD}\right]^{-\epsilon /2}.
	\end{equation}
    This tells us that the localization of polymers in response to activity is governed by the exponent $\epsilon$, which encapsulates the competition between the effective drift, which causes the directed motion toward high activity regions, and effective diffusion, which tends to displace the polymer from its residing place. In particular, it is the sign of $\epsilon$ (or $S_2 - S_1$) that determines whether a polymer prefers to localize in high or low activity regions, with $\epsilon<0$ leading to preferential accumulation in high activity regions or chemotaxis. 

    We note that the dynamic equations for the center of mass of the active polymer is determined by the dimensionless scaling or activity parameter 
    \begin{equation}\label{eqn:kappa}
        \kappa = \tau\gamma = \frac{\tau}{\tau_m}\gg 1\;\;,
    \end{equation}
which denotes the ratio of persistence time of the direction of the activity with respect to the diffusive monomer relaxation time, $\tau_m=1/\gamma$ and should be always much larger then unity. The characteristic time scales can be then expressed units of $\tau$.
	
	Fig. \ref{fig:steadystateprofiles} shows the steady state density profiles of various configurations of polymers for chain lengths $N=6$ and $N=18$, in an environment with a sinusoidal fuel concentration. Let us first consider the density distribution of polymers with a single active monomer compared to that with all monomers active. 
 A polymer with only an end active
monomer shows stronger accumulation in high activity regions compared to a polymer composed uniquely by active units. However, if the active monomer is located in the interior of the chain, the polymer’s localization is weaker than an all-active polymer. This can be explained by the value of the exponent $\epsilon$, which as mentioned before determines the location and effectiveness of the preferential accumulation. Precisely, the more negative is the value of $\epsilon$, the more efficient is the accumulation of polymer in high activity regions. Therefore, by evaluating the values of $\epsilon$ (Eqn \eqref{eqn:epsilon}), we see that $|\epsilon_{end}| > |\epsilon_{all}| > |\epsilon_{interior}|$ which supports the results presented in Fig. \ref{fig:steadystateprofiles}, which holds good independent of the polymer length, as can be seen from the results for $N=18$ in Fig. \ref{fig:steadystateprofiles}. 

The limit $N \gg 1$ can be considered analytically for selected configurations of chains with one active monomer. In this case the summations in Eqs.(\ref{eqn:S}) can be transformed into integrals leading to closed analytic expressions as shown in the supplementary material. In particular we obtain for $\epsilon$:
\begin{equation}\label{eqn:epsilon_relations_gen}
    |\epsilon_{end}| = \frac{N}{4\sqrt{\kappa}} -1 \simeq 2 |\epsilon_{mid}|\;,
\end{equation}
where "mid" denotes the center monomer.  Here, the case $\kappa \gg 1$ is considered. Similarly, we can solve the case of the all-active chain with the result: $\epsilon_{all}\simeq\epsilon_{mid}$. For chemotactic behavior ($\epsilon <0$) we have to further assume $N^2 \gg \kappa$, or $\tau \ll \tau_m N^2 =\tau_R$, which means that the active persistence time should be much smaller than the diffusive relaxation time, or Rouse time $\tau_R$ , of the whole chain. This result is in agreement with the previous findings for all-active chains\cite{muzzeddu2024migration}.
	
Using the symmetry of the eigenfunctions of the connectivity matrix for the linear chain (see supplementary material) we draw some further general conclusions about the role of the position of the active monomers inside the chain: Consider polymers that have a symmetric distribution of active monomers along the chain, e.g. a polymer with both terminal monomers active, and their corresponding antisymmetric polymers in which the active monomers are present in only one of half of the chain, e.g for the above mentioned case is a polymer chain with only one end monomer active. Using the fact that the absolute values of the elements of the eigenvectors form a palindromic set, we can show that the values of $S_1$ and $S_2$ (Eqn \eqref{eqn:S}) for a symmetric polymer are twice of those of its corresponding antisymmetric polymer. This leads to the same epsilon value (Eqn \eqref{eqn:epsilon}) for both cases and as a result, their steady state densities differ only marginally due to the difference in the pre-factor in Eqn. \eqref{eqn:steadystate} , as can be seen in Fig.\ref{fig:steadystateprofiles}. This further emphasizes the fact that accumulative behaviour is largely determined by  $\epsilon$ and shows that such a pair of symmetric-antisymmetric polymers have similar localization with respect to inhomogeneous activity.
	
	\begin{figure}[h]
		\centering
		\includegraphics[width=0.47\textwidth]{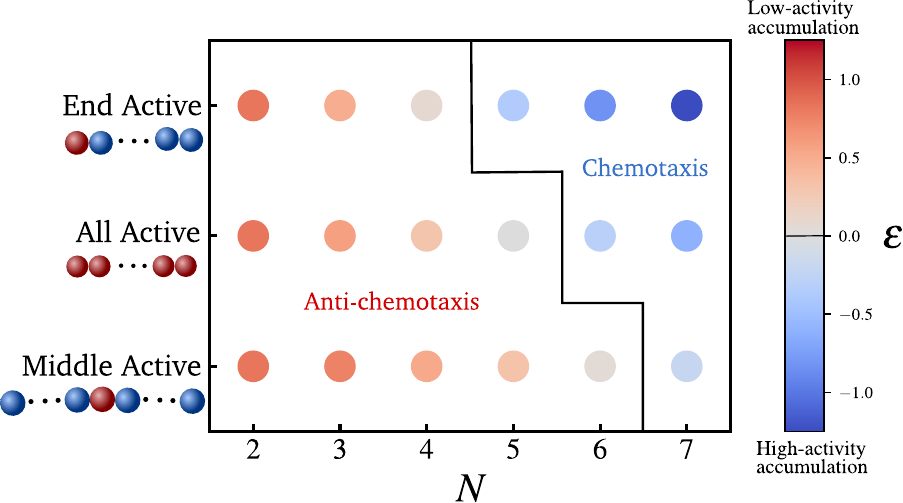}
		\caption{\label{fig:chemostatediagram} Scatter plot for $\epsilon$ which indicates the location and degree of preferential accumulation for three distinct configuration of polymers of various chain length. Values of $\epsilon$ obatined analytically for the paramters:  $k = 12$, $D_r=5$, and $\mu = 1$. As mentioned in the main text,  $\epsilon < 0$ on the colorbar indicates accumulation in high activity regions, and those greater than zero represent low-activity accumulation.}
	\end{figure}
	
	Therefore, limiting our focus to polymers with just one or all monomers active, we construct in Fig. \ref{fig:chemostatediagram} a state diagram of the accumulation behavior of hybrid polymers. In particular, we consider polymers with different chain lengths $N$ and three different configuration: active end monomer active, active central monomer active, and all monomers active. We observe that the switch from accumulation in lower activity regions to higher activity regions, for a given set of parameter values ($\zeta$,$\mu$,$\tau$), can be achieved not only by increasing the chain length or varying the connectivity matrix as reported in Ref. \cite{muzzeddu2024migration}, but also by altering the number and positions of the active monomers within the polymer. Note that when all the monomers are active ($\alpha_i = 1$ for all $i$), the expression for the steady state density simplifies to the one presented in \cite{muzzeddu2024migration} for polymers made up of Active Ornstein Uhlenbeck particles (AOUP's)\cite{martin2021statistical, caprini2018active}. This illustrates that the accumulative behaviour of polymers is not dependent on the specific model of active particles considered.

	\begin{figure}[h]
		\centering
		\includegraphics[width=0.45\textwidth]{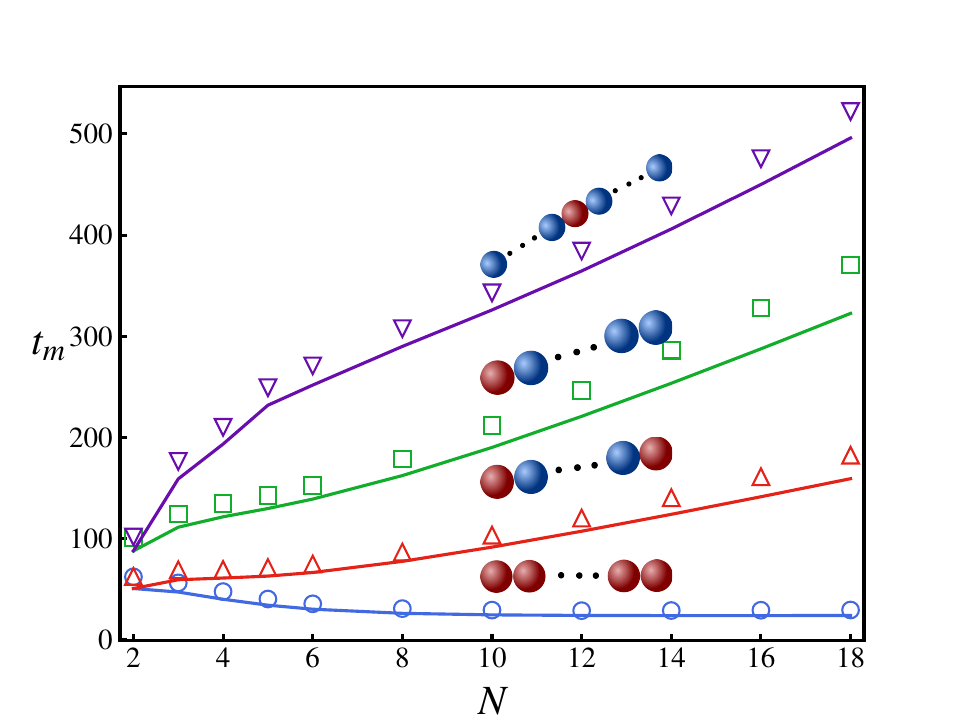}
		\caption{\label{fig:mfpt} The mean first passage time taken by a polymer to travel to the most active location starting from $x=L/2$ vs the chain length of the polymer for four different configurations (other parameters are same as those specified in Fig. \ref{fig:steadystateprofiles}). The solid lines are theoretical predictions while the symbols are from Langevin dynamics simulations.}
	\end{figure}

    Though Eqn. \eqref{eqn:steadystate} gives us the steady state accumulation of various polymer chains, it does not give any insights into their dynamic behaviour, for example how fast do they get to the regions of high fuel concentration. In this direction, we solve for the mean first passage time (MFPT) $T(x)$, $x$ being the initial position of the polymer's center of mass, the equation for which is obtained using the coarse-grained Fokker-Planck equation as \cite{gardiner1985handbook}-
    \begin{equation}
    \left( \left(1 - \frac{\epsilon}{2} \right) \partial_x \mathcal{D} \right) \partial_x T(x) + \mathcal{D}\partial^2_x T(x) = -1.
    \end{equation} 
    Setting $t_m = T(50)$ for a box of length $L=100$, we calculate the average time taken by a polymer to reach the location of highest activity by imposing absorbing boundary conditions \cite{gardiner1985handbook}. The results for the mean first passage time for various polymer configurations are presented in Fig. \ref{fig:mfpt}. For a polymer with all monomers active, the MFPT decreases as we increase the number of monomers, becoming nearly independent of the chain length for $N>10$. However, other polymers with one or two monomers active take longer to reach the region of highest activity. As can be evinced from Fig.~\ref{fig:mfpt}, the MFPT of polymers with a single active monomer varies depending on the position of the latter along the chain, reaching the minimum value when the active monomer is an end monomer. Furthermore, in contrast to the previously discussed accumulation behavior, Fig.~\ref{fig:mfpt} shows that introducing a further active monomer in a symmetric position compared to a preexisting one (as shown by the green and red solid lines in Fig.~\ref{fig:mfpt}) has a tangible impact on the MFPT. In particular, the configuration with two monomers (red lines) is much faster compared to the other (green line).
    These results suggest that increasing the number of active units within a polymer decreases its MFPT.
	
\begin{figure}[t]
		\centering
		\includegraphics[width=0.40\textwidth]{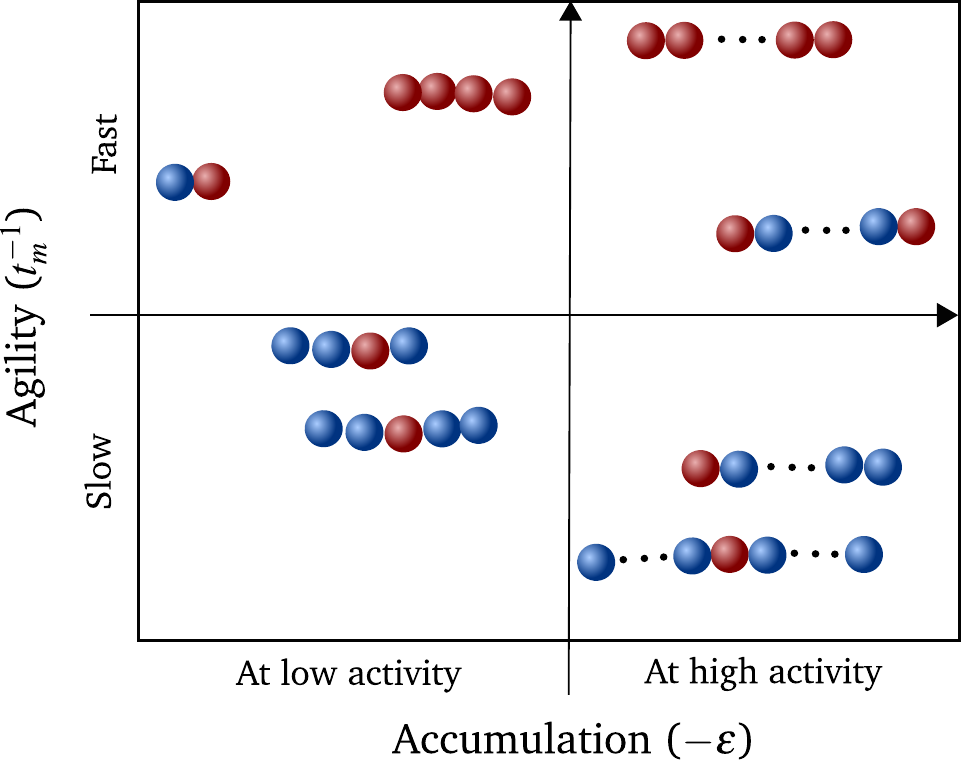}
		\caption{\label{fig:mfptstatediagram} A qualitative state diagram that illustrates the two studied properties of various polymers: (i) Location of accumulation on the x-axis - by taking the negative of $\epsilon$ and (ii) Mean first passage time of polymers to reach the location of highest activity $t_m$.}
	\end{figure}

	Constructing a qualitative state diagram that encapsulates both the agility (defined as the inverse of MFPT) and the preferential accumulation behavior of polymers in Fig. \ref{fig:mfptstatediagram}, we get a comprehensive picture regarding hybrid polymers in chemical gradients. There seems to be no correlation between preferential accumulation and the mean first passage time of the polymers. We can therefore have polymers ranging from those that accumulate in high activity regions but are slow getting there (like long polymers ($N>10$) with a sole interior monomer active) to those that are fast in getting to the high active regions but do not prefer to localize there (like relatively short polymers ($N=4$) with all monomers active). 

From our studies, we can therefore suggest strategies for polymerization that leads to (i) maximum accumulation and (ii) fastest motion, for a given $N$. A polymer with both terminal monomers active would result in the former while the one with all units active would be the fastest. This information may shed light onto the evolutionary development of important biopolymers like actin \cite{straub1943actin,khaitlina2014intracellular} and tubulin \cite{hamel1984separation}, even though the form of activity at these length scales are different. The best strategy required for the function of the biopolymer can still be constructed by altering the number and position of the active units within the polymer. Similar self-localization of symmetric and corresponding anti-symmetric polymers can also be an indicator of introduction of asymmetry in biopolymers during evolution to yield an energy efficient directed transport.

In the current work, we have focused on polymers of fixed lengths. The incorporation of polymerization and depolymerization into a dynamical model is left for a future work. Our findings can also be coupled with existing studies on polymerization induced phase separation to study the formation and behaviour of active cell organelles \cite{weber2019physics} in future works.

\section*{Conflicts of interest}
There are no conflicts to declare.

\section*{Data availability}
The code and primary data can be found at \url{https://zenodo.org/records/14168683} with  DOI: 10.5281/zenodo.14168683.

\section*{Acknowledgements}
A.S. acknowledges support by the Deutsche Forschungsgemeinschaft (DFG)  within Project No. SH 1275/5-1. J.U.S. was supported by the Deutsche Forschungsgemeinschaft (DFG,
German Research Foundation) under Germany's Excellence Strategy --
EXC 2068 -- 390729961-- Cluster of Excellence Physics of Life of
TU Dresden.

\bibliographystyle{references1}
\bibliography{references}
\onecolumngrid
\pagenumbering{Roman}
\renewcommand{\thefigure}{SI~\arabic{figure}}
\setcounter{figure}{0}
\renewcommand{\thetable}{SI~\arabic{table}}
\setcounter{table}{0}
\renewcommand{\theequation}{SI~\arabic{equation}}
\setcounter{equation}{0}
\vspace{1.5in}

\begin{center}
\textbf{\large Supplemental Material: Transport of molecules via polymerization in chemical gradients}
\end{center}

\section{The Langevin equations, Rouse modes and the Fokker-Planck equation}
The set of overdamped Langevin equations describing the motion of the polymer is
	\begin{align}
		\label{eqn:langevin}
		\bm{\dot{X}}_i(t) &= -\mu \nabla_{\bm{X}_i}\mathcal{H} + \mu \alpha_i f_s(\bm{X}_i)\bm{p}_i + \bm{\xi}_i(t),\\
		\label{eqn:langevin_orientation}
		\bm{\dot{p}}_i(t) &= \bm{p}_i \times \bm{\eta}_i(t),
	\end{align} 
	where $i = \lbrace0,1,2\dots N-1\rbrace$, $N$ being the number of monomers in the polymer chain, $\mu$ is the mobility of the monomers, $f_s$ is the active force, which is a function of the spatial coordinates and it acts along the orientation vector $\lbrace\bm{p}_i\rbrace$, which evolves in time due to rotational diffusion. The nature of monomers are taken care by the parameters $\lbrace \alpha_i$ $\rbrace$, with $\alpha_i = 1$ and $0$ corresponding to active and passive monomers, respectively.  $\lbrace \bm{\xi}_i (t)\rbrace$ and $\lbrace \bm{\eta}_i (t)\rbrace$ are zero-mean white Gaussian noises such that
	\begin{align}
		\langle\bm{\xi}_i(t) \otimes \bm{\xi}_j(s)\rangle &= 2D\bm{I}\delta_{ij}\delta (t-s),\\
		\langle\bm{\eta}_i(t) \otimes \bm{\eta}_j(s)\rangle &= 2D_r\bm{I}\delta_{ij}\delta (t-s),
	\end{align}
	where $D$ is the thermal diffusivity and $D_r$ is the rotational diffusivity, and $\bm{I}$ is the $d \times d$ identity matrix ($d$ denotes the number of dimensions). The interactions between monomers 
 are modeled by a harmonic potential and are governed by the Hamiltonian $\mathcal{H}$,
	\begin{equation}
		\mathcal{H} = \frac{\zeta}{2}\sum_{ij}M_{ij}\bm{X}_i\cdot\bm{X}_j,
	\end{equation}
	where $M_{ij}$ is the connectivity matrix of the polymer and $\zeta$ is the spring constant.
	
	As done frequently in polymer physics, we obtain the Rouse modes from the physical coordinates of the monomers via the linear transformation 
	\begin{equation}
		\bm{\chi}_i = \sum_{j} \varphi_{ij} \bm{X}_j,
	\end{equation}
	where $\varphi_{ij}$ is a matrix that diagonalizes the connectivity matrix $\bm{M}$ such that $\sum_{jk} \varphi_{ij}M_{jk}\varphi^{-1}_{kl} = \frac{\gamma_i}{\gamma}\delta_{il}$, $\gamma= \mu\zeta$ being an inverse timescale due to the spring relaxation. Applying this linear transformation to Eq. \eqref{eqn:langevin}, we get the time evolution of the Rouse modes:
	\begin{equation}
		\label{eqn:rouse_dynamics}
		\bm{\dot{\chi}}_i = \gamma_i \bm{\chi}_i + \sum_j \varphi_{ij}\alpha_jv(\bm{X}_j)\bm{p}_j + \tilde{\bm{\xi}}_i(t),
	\end{equation}
	where $v(\bm{X}_j) = \mu f_s(\bm{X}_j)$ is the swim speed of the monomers and $\lbrace\tilde{\bm{\xi}}_i(t)\rbrace$ are  Gaussian white noises with the same statistical properties as $\lbrace\bm{\xi}_i(t)\rbrace$.
	
	The corresponding Fokker-Planck equation for \eqref{eqn:langevin_orientation} and \eqref{eqn:rouse_dynamics} is given by 
	\begin{equation}
		\label{eqn:FPE}
		\partial_t \mathcal{P} = (\mathcal{L}_0 + \mathcal{L}_a + \mathcal{L}_{\bm{p}})\mathcal{P},
	\end{equation}
	where $\mathcal{P}(\lbrace \bm{\chi} \rbrace, \lbrace \bm{p} \rbrace, t)$ is the joint probability density. The operators in \eqref{eqn:FPE} are defined as
	\begin{equation}
        \begin{split}
		\mathcal{L}_0 &\equiv \sum_{i=0}^{N-1} \nabla_i \cdot \left[\gamma_i \bm{\chi}_i + D\nabla_i \right], \\
		\mathcal{L}_a &\equiv - \sum_{i=0}^{N-1} \nabla_i \cdot \left[\sum_j \varphi_{ij}\alpha_j v(\bm{X}_j)\bm{p}_j \right], \\
		\mathcal{L}_{\bm{p}} &\equiv \sum_{i=0}^{N-1} D_r\tilde{\nabla}_i^2,
        \end{split}
	\end{equation}
	where $\nabla_i \equiv \nabla_{\bm{\chi}_i}$ and $\tilde{\nabla}_i \equiv \nabla_{\bm{p}_i}$.

 \section{Coarse-graining of the Fokker-Planck Equation}
	
	Since we are only interested in the spatial distribution of these polymers, we look for a probability density $\rho$ that is only a function of the 
 center of mass of the polymer $\bm{X}_{\rm{COM}} = \bm{\chi}_0/\sqrt{N}$. To obtain this we carry out two steps of coarse graining : (i) integrating out all the orientation vectors to obtain a marginal density 
	\begin{equation}
		\varrho = \left(\lbrace \bm \chi \rbrace, t\right) = \int \prod_i d\bm{p}_i P(\lbrace \bm \chi \rbrace, \lbrace \bm p \rbrace, t),
	\end{equation}  
	and (ii) integrating out the rest of the Rouse modes to obtain the probability distribution function $\rho_0$
	\begin{equation}
		\label{eqn:marginalcom}
		\rho_0 (\bm{\chi}_0, t) = \int \prod_{i\neq 0}d\bm{\chi}_i \varrho(\lbrace \bm \chi \rbrace, t).
	\end{equation}

To carry out the first step of coarse-graining, we expand the joint probability distribution function using spherical harmonics:
\begin{equation}
P(\lbrace \bm \chi \rbrace, \lbrace \bm p \rbrace, t) = \frac{1}{\Omega_d^N}\left(\phi + \sum_i \bm{\sigma}_i\cdot\bm{p}_i + \sum_{i\neq j} \bm{\sigma}_{ij}:\bm{p}_i\bm{p}_j + \sum_i \bm{\omega}_i:\left(\bm{p}_i\bm{p}_i - \bm{I}/d \right)  + \Theta \right),
\end{equation}
constructed using the eigenfunctions $1$, $\lbrace{\bm{p}_i\rbrace}$, $\lbrace{\bm{p}_i\bm{p}_i - \bm{I}/d \rbrace}$, with the eigenvalues $0$, $-(d-1)$, and $-2d$, respectively. $\phi$, $\bm{\sigma}_i$, $\bm{\sigma}_{ij}$, and $\bm{\omega}_i$ are the modes of the expansion and are functions of $\lbrace \bm{\chi} \rbrace, t$. $\phi$ is the marginal density $\varrho$, $\lbrace \bm{\sigma}_i \rbrace$ are related to the average orientations, while $\bm{\omega}$ contains information about the nematic order. The function $\Theta$ takes into account all the higher order modes and eigenfunctions. We are interested the steady state distribution of the density in small gradients, and hence we ignore the $\lbrace \bm{\omega}_i \rbrace$, $\lbrace \bm{\sigma}_{ij}\rbrace$, and $\Theta$ terms that lead to $\mathcal{O}(\nabla^2)$ terms after coarse-graining. More details about this can be found in Ref. \cite{vuijk2021chemotaxis, muzzeddu2022active}. The truncated expansion for $P$ thus reads -
\begin{equation}
    P(\lbrace \bm \chi \rbrace, \lbrace \bm p \rbrace, t) = \frac{1}{\Omega_d^N}\left(\phi + \sum_i \bm{\sigma}_i\cdot\bm{p}_i \right).
\end{equation}

Before integrating out the orientation vectors, we define the scalar product
\begin{equation}
\langle f,g \rangle = \int \prod_i d\bm{p}_i f(\lbrace \bm{p} \rbrace) g(\lbrace \bm{p} \rbrace),
\end{equation}
and list out the following identities that will be useful while integrating -
\begin{align}
\langle 1,\mathcal{P} \rangle &= \phi = \varrho, \\
\langle \bm{p}_j,\mathcal{P} \rangle &= \frac{
1}{d}\bm{\sigma}_j,\\
\langle \bm{p}_j \bm{p}_j,\mathcal{P} \rangle &= \frac{1}{d}\phi\bm{I},\\
\langle \bm{p}_j \bm{p}_k,\mathcal{P} \rangle &= 0, \\
\langle 1,\sum_i\tilde{\nabla}^2_i\mathcal{P} \rangle &= 0, \\
\langle \bm{p}_j,\sum_i\tilde{\nabla}^2_i\mathcal{P} \rangle &= -\frac{
1}{d}\bm{\sigma}_j,\\
\langle \bm{p}_j \bm{p}_k,\sum_i\tilde{\nabla}^2_i\mathcal{P} \rangle &= 0.
\end{align}

Taking the scalar product of \eqref{eqn:FPE} with $1$ and $\bm{p}_j$ gives us the coarse grained dynamics of the first two modes of the expansion-
\begin{align}
\partial_t \varrho &= \sum_i \nabla_i \left[\gamma_i \bm{\chi}_i\varrho + D\nabla_i\varrho - \sum_j\varphi_{ij}\alpha_j v(\bm{X}_j)\frac{\bm{\sigma}_j}{d} \right],\\
\partial_t \bm{\sigma}_j &= -\tau^{-1}\bm{\sigma}_j + \sum_l \nabla_l\cdot \left[\gamma_l \bm{\chi}_l \bm{\sigma}_j + D\nabla_l \bm{\sigma}_j - \varphi_{lj}\alpha_jv(\bm{X}_j)\varrho  \right],
\label{eqn:sigmajcomplete}
\end{align}
where $\tau^{-1} = (d-1)D_r$. The equation for the dynamics of $\varrho$ can be written as a continuity equation
\begin{equation}
\label{eqn:conserved}
\partial_t \varrho = - \sum_i \nabla_i \cdot \bm{J}_i,
\end{equation}
where the fluxes are given by
\begin{equation}
\bm{J}_i = -\gamma_i\bm{\chi}_i\varrho + \sum_j\varphi_{ij}\alpha_j v(\bm{X}_j)\frac{\bm{\sigma_j}}{d} - D\nabla_i\varrho.
\end{equation}

Proceeding with the second step of coarse-graining, using the definition in \eqref{eqn:marginalcom}, we get
\begin{equation}
\label{eqn:coarsegrainedfpe}
\partial_t \rho_0 = - \nabla_0\cdot \bm{\mathcal{J}}_0,
\end{equation}
where
\begin{equation}
\label{eqn:comflux}
\bm{\mathcal{J}}_0 =  \sum_j\frac{\varphi_{0j}}{d}\int\prod_{h\neq 0}d\bm{\chi}_h \alpha_j v(\bm{X}_j)\bm{\sigma}_j - D\nabla_0\rho_0,
\end{equation}
where we have used the fact that $\gamma_0 = 0$ (see Sec. \ref{sec:rouse} for more details).

Since $\varrho$ satisfies a conserved equation (Eq. \eqref{eqn:conserved}), it is the slowest mode \cite{vuijk2021chemotaxis}. Therefore the time derivative terms in the time-evolution equation for $\lbrace \bm{\sigma}_j \rbrace$'s can be neglected in \eqref{eqn:sigmajcomplete} as they decay on a much faster timescale compared to $\varrho$. Thus, we get -
\begin{equation}
\label{eqn:sigmaj}
\bm{\sigma}_j = \sum_l \tau\nabla_l\cdot \left[\gamma_l \bm{\chi}_l \bm{\sigma}_j + D\nabla_l \bm{\sigma}_j - \varphi_{lj}\alpha_jv(\bm{X}_j)\varrho \right].
\end{equation}
Plugging it into the eqn \eqref{eqn:comflux}, and isolating the contribution by the active term we get 
\begin{equation}
\begin{split}
\bm{\mathcal{J}}_0 ^{act} =& \sum_j\frac{\varphi_{0j}}{d}\int\prod_{h\neq 0}d\bm{\chi}_h \alpha_j v(\bm{X}_j)\bm{\sigma}_j, \\
=& \underbrace{-\sum_{j} \frac{\varphi_{0j} \tau}{d} \int\prod_{h\neq 0}d\bm{\chi}_h \alpha_j v(\bm{X}_j)\sum_l\nabla_l \cdot \left[\varphi_{lj}\alpha_jv(\bm{X}_j)\varrho \right]}_{\bm{\mathcal{J}}_0 ^{act,1}} \\
& + \underbrace{\sum_{j} \frac{\varphi_{0j} \tau}{d} \int\prod_{h\neq 0}d\bm{\chi}_h \alpha_j v(\bm{X}_j)\sum_l\nabla_l \cdot \left[\gamma_l \bm{\chi}_l \bm{\sigma}_j \right]}_{\bm{\mathcal{J}}_0 ^{act,2}},
\end{split}
\end{equation}
where the $\nabla_l \bm{\sigma}_j$ term in eqn. \eqref{eqn:sigmaj} contributes a term of order $\mathcal{O}(\nabla_0^2)$ (see Ref~\cite{muzzeddu2024migration}), which can be neglected for small gradients.

To calculate $\bm{\mathcal{J}}_0 ^{act,1}$, we split the summation over $l$ into terms with $l=0$ and $l \neq 0$. The latter gives 
\begin{align}
&-\sum_{j} \frac{\varphi_{0j} \tau}{d} \int\prod_{h\neq 0}d\bm{\chi}_h \alpha_j v(\bm{X}_j)\sum_{l\neq 0}\nabla_l \cdot \left[\varphi_{lj}\alpha_jv(\bm{X}_j)\varrho \right], 
\\
&= \frac{\tau}{2d}\sum_{j, l\neq 0} \varphi_{0j} \int\prod_{h\neq 0}d\bm{\chi}_h \sqrt{N}\alpha_j^2 \varphi_{lj}\varphi_{lj}\varrho \nabla_0 \left(v^2(\bm{X}_j) \right),
\\
&= \frac{\tau}{2d}\underbrace{\left[\sum_{j, l\neq 0} \alpha_j^2 \varphi_{lj}^2\right]}_{S_1'}  \rho_0 \nabla_0 \left(v^2\left(\frac{\bm{\chi}_0}{\sqrt{N}}\right) \right) + \mathcal{O}(\nabla_0^2),
\end{align}
where we have used integration by  parts and
\begin{align}
\nabla_l v(\bm{X}_j) &= \sqrt{N}\varphi_{lj}\nabla_0 v(\bm{X}_j),\\
\nabla_0 v^2(\bm{X}_j) &= \nabla_0 v^2(\varphi_{lj}\bm{\chi}_l) = \nabla_0 v^2(\varphi_{0j}\bm{\chi}_0) + \mathcal{O}(\nabla_0^2),\\
\varphi_{0j} &= \frac{1}{\sqrt{N}} . \label{eqn:phi0j} 
\end{align}
The $l=0$ term is simplified as
\begin{align}
&-\sum_{j} \frac{\varphi_{0j} \tau}{d} \int\prod_{h\neq 0}d\bm{\chi}_h \alpha_j v(\bm{X}_j)\nabla_0 \cdot \left[\varphi_{0j}\alpha_jv(\bm{X}_j)\varrho \right],
\\
&= -\frac{\tau}{2d}\underbrace{\left[\sum_j \alpha_j^2\varphi_{0j}^2 \right]}_{S_2}\rho_0 \nabla_0\left(v^2\left(\frac{\bm{\chi}_0}{\sqrt{N}} \right)\right) - \frac{\tau}{d}\left[\sum_j \alpha_j^2\varphi_{0j}^2 \right] v^2\left(\frac{\bm{\chi}_0}{\sqrt{N}} \right)\nabla_0 \rho_0  + \mathcal{O}(\nabla_0^2),
\end{align}
where the summation $S_2$ can be simplified using Eqn. \eqref{eqn:phi0j} to obtain -
\begin{equation}
    S_2 = \frac{1}{N}\sum_{j=0}^{N-1} \alpha_j^2 = \frac{1}{N}\sum_{j=0}^{N-1} \alpha_j.
\end{equation}
Combining the two terms we get 
\begin{equation}
\bm{\mathcal{J}}_0 ^{act,1} = \frac{\tau}{2d}(S_1' - S_2)\rho_0 \nabla_0\left(v^2\left(\frac{\bm{\chi}_0}{\sqrt{N}} \right)\right) - \frac{\tau}{d}S_2 v^2\left(\frac{\bm{\chi}_0}{\sqrt{N}} \right)\nabla_0 \rho_0  + \mathcal{O}(\nabla_0^2).
\end{equation}

Representing $\bm{\mathcal{J}}_0^{act,2}$ as
\begin{equation}
\label{eqn:activeflux2}
\bm{\mathcal{J}}_0^{act,2} = \sum_l \bm{\mathcal{I}}_l,
\end{equation}
we have
\begin{align}
\bm{\mathcal{I}}_l &= \sum_j\frac{\varphi_{0j} \tau}{d} \int\prod_{h\neq 0}d\bm{\chi}_h \alpha_j v(\bm{X}_j)\left[ \nabla_l \cdot \left(\gamma_l \bm{\chi}_l \bm{\sigma}_j \right)\right],\\
&= \sum_j -\frac{\varphi_{0j} \tau}{d} \int\prod_{h\neq 0}d\bm{\chi}_h  \left[ \nabla_l \left( \alpha_j v(\bm{X}_j) \right)\right] \cdot \gamma_l \bm{\chi}_l \bm{\sigma}_j.
\end{align}
Substituting the expression for $\bm{\sigma}_j$ using eqn. \eqref{eqn:sigmaj} -
\begin{equation}
\begin{split}
\bm{\mathcal{I}}_l = \sum_j \frac{\varphi_{0j} \tau^2}{d} \sum_m\int\prod_{h\neq 0}d\bm{\chi}_h  \left[ \nabla_l \left( \alpha_j v(\bm{X}_j) \right)\right] \cdot \gamma_l \bm{\chi}_l \left[\nabla_m \left(\varphi_{mj}\alpha_j v(\bm{X}_j)\varrho\right) - \nabla_m\cdot\left(\gamma_m\bm{\chi}_m\bm{\sigma}_j \right)\right] + \mathcal{O}(\nabla_0^2).
\end{split}
\end{equation}
The first term is simplified as
\begin{align}
&\sum_j \frac{\varphi_{0j} \tau^2}{d} \sum_m\int\prod_{h\neq 0}d\bm{\chi}_h  \left[ \nabla_l \left( \alpha_j v(\bm{X}_j) \right)\right] \cdot \gamma_l \bm{\chi}_l \left[\nabla_m \left(\varphi_{mj}\alpha_j v(\bm{X}_j)\varrho\right)\right],
\\ 
&=-\sum_j \frac{\varphi_{0j} \tau^2}{d} \int\prod_{h\neq 0}d\bm{\chi}_h \left[ \nabla_l \left( \alpha_j v(\bm{X}_j) \right)\right]\gamma_l  \varphi_{lj}\alpha_j v(\bm{X}_j)\varrho + \mathcal{O}(\nabla_0^2),
\\
&=-\frac{\gamma_l \tau^2}{2d} \sqrt{N} \left[\sum_j \varphi_{0j}\varphi_{lj} \varphi_{lj} \alpha_j^2\right] \rho_0 \nabla_0 \left( v^2\left(\frac{\bm{\chi}_0}{\sqrt{N}} \right)\right)  + \mathcal{O}(\nabla_0^2),
\\
&=-\frac{\gamma_l \tau^2}{2d} \left[\sum_j \varphi_{lj}^2 \alpha_j^2\right] \rho_0 \nabla_0 \left(v^2 \left(\frac{\bm{\chi}_0}{\sqrt{N}} \right)\right)  + \mathcal{O}(\nabla_0^2).
\end{align}
The second term simplifies to
\begin{align}
&-\sum_j \frac{\varphi_{0j} \tau^2}{d} \sum_m\int\prod_{h\neq 0}d\bm{\chi}_h  \left[ \nabla_l \left( \alpha_j v(\bm{X}_j) \right)\right] \cdot \gamma_l \bm{\chi}_l \left[\nabla_m\cdot\left(\gamma_m\bm{\chi}_m\bm{\sigma}_j \right)\right],\\
& =\sum_j \frac{\varphi_{0j} \tau^2}{d} \int\prod_{h\neq 0}d\bm{\chi}_h  \left[ \nabla_l \left( \alpha_j v(\bm{X}_j) \right)\right] \cdot \gamma_l  \gamma_l\bm{\chi}_l\bm{\sigma}_j + \mathcal{O}(\nabla_0^2),\\
& = -\tau\gamma_l \bm{\mathcal{I}}_l + \mathcal{O}(\nabla_0^2).
\end{align}
Combining the two terms and substituting in eqn. \eqref{eqn:activeflux2}, we get
\begin{equation}
\bm{\mathcal{J}}_0^{act,2} = -\frac{\tau}{2d} \underbrace{\left[\sum_{lj}\frac{\tau\gamma_l}{1+\tau\gamma_l} \varphi_{lj}^2 \alpha_j^2\right]}_{S_1^*} \rho_0 \nabla_0 \left(v^2\left(\frac{\bm{\chi}_0}{\sqrt{N}} \right)\right).
\end{equation}

Therefore the complete expression for the flux $\bm{\mathcal{J}}_0$ (Eqn. \eqref{eqn:comflux}) is 
\begin{equation}
\label{eqn:J0}
\bm{\mathcal{J}}_0 = \frac{\tau}{2d}(S_1 - S_2)\rho_0 \nabla_0\left(v^2\left(\frac{\bm{\chi}_0}{\sqrt{N}} \right)\right) - \left[D + \frac{\tau}{d}S_2v^2\left(\frac{\bm{\chi}_0}{\sqrt{N}}\right)\right]\nabla_0 \rho_0,
\end{equation}
where $S_1 = S_1' - S_1^*$,
\begin{equation}
\label{eqn:S1}
    S_1 = \sum_{l=1,j=0}^{N-1}\frac{1}{1+\tau\gamma_l} \varphi_{lj}^2 \alpha_j^2.
\end{equation}

We now change the variable to $\bm{X}_{\rm{COM}} = \bm{\chi}_0/\sqrt{N}$ and rewrite the coarse-grained Fokker-Planck equation in terms of the function $\rho(\bm{X}_{\rm{COM}},t) = \rho_0(\chi_0,t)$. Dropping subscripts and the arguments,  the flux in eqn. \eqref{eqn:J0} can be written as 
\begin{equation}
\bm{\mathcal{J}} = \rho\bm{\mathcal{V}} - \nabla(\mathcal{D}\rho),
\end{equation} 
where the gradients are with respect to the centre of mass coordinates and $\bm{\mathcal{V}}$ and $\mathcal{D}$ are the effective drift and effective diffusion coefficient, respectively, given by
\begin{equation}
    \label{eqn:coefficients}
    \begin{split}
        \bm{\mathcal{V}}(\bm{X}_{\rm{COM}}) &= \frac{\tau}{dN}\left(\frac{S_1 + S_2}{2}\right)\nabla\left(v^2\left(\bm{X}_{\rm{COM}} \right)\right),\\
        \mathcal{D}(\bm{X}_{\rm{COM}}) &= \frac{1}{N}\left(D + \frac{\tau}{d}S_2v^2\left(\bm{X}_{\rm{COM}}\right)\right).
    \end{split}
\end{equation}
The two terms are related by $\bm{\mathcal{V}} =(1  -\frac{\epsilon}{2})\nabla \mathcal{D}$, where 
    \begin{equation}
    \label{eqn:epsilon}
		\epsilon = \frac{S_2 - S_1}{S_2},
	\end{equation}

\section{Eigenvectors and eigenvalues for a linear chain}
\label{sec:rouse}
The eigenvector matrix for a linear chain can be obtained as \cite{doi1988theory}-
\begin{equation}
  \setlength{\arraycolsep}{0pt}
  \varphi_{lj} = \left\{ \begin{array}{ l l }
    \sqrt{\frac{1}{N}},  \;\;\;\; &(l=0) \\
    \sqrt{\frac{2}{N}}\cos \left(\frac{l\pi}{N}\left(j + \frac{1}{2} \right) \right), \;\;\; &(l\neq 0)
  \end{array} \right.
\end{equation}
and the normalized eigenvalues are given by -
\begin{equation}
    \gamma_l = 4 \gamma \sin ^2 \left(\frac{l\pi}{2N}\right).
\end{equation}
It can be easily verified that $|\varphi_{lj}|$ or $\varphi_{lj}^2$ is invariant for $j \to N-1-j$. Consider a polymer whose active monomers are distributed symmetrically along the chain (ex - both end monomers active). It can now be shown, using the invariance of $\phi_{lj}^2$ under the aforementioned transformation, that the values of $S_1$ (Eq. \eqref{eqn:S1}) and $S_2$ (the fraction of active monomers in the polymer) are twice their values for the corresponding case when only one half of the polymer has active monomers (ex - only one end monomer active). This gives us the same value of epsilon (Eq. \eqref{eqn:epsilon}) for both cases.

\begin{center}
    \textbf{Comparison between end monomer active and central monomer active}
\end{center}
For a polymer with one end monomer active, we have:
\begin{align}
    S_1^{end} &= \sum_{l=1}^{N-1} \frac{1}{1+\tau \gamma_l} \phi_{l0}^2 = \frac{2}{N} \sum_{l=1}^{N-1} \frac{\cos ^2 \left( \frac{l\pi}{2N} \right) }{1+ 4\tau\gamma \sin ^2 \left(\frac{l\pi}{2N}\right)},
\end{align}
while for a polymer with the central monomer active:
\begin{align}
    S_1^{mid} &= \sum_{l=1}^{N-1} \frac{1}{1+\tau \gamma_l} \phi_{l\frac{N}{2}}^2 = \frac{2}{N} \sum_{l=1}^{N-1} \frac{ \cos ^2 \left( \frac{l(N+1)\pi}{2N} \right)}{1 + 4\tau\gamma \sin ^2 \left(\frac{l\pi}{2N}\right)}\;.
\end{align}
For both cases
\begin{equation}
    S_2 = \frac{1}{N}\;.
\end{equation}
In the limit of $N\gg 1$ we can rewrite $S_1^{mid}$ as

\begin{equation}
    S_1^{mid} = \frac{2}{N} \sum_{l=2,4,...}^{N-1} \frac{1}{1+\tau\gamma\sin^2 \left(\frac{l\pi}{2N}\right)}\simeq \frac{1}{N} \sum_{l=1}^{N-1} \frac{1}{1+\tau\gamma\sin^2 \left(\frac{l\pi}{2N}\right)}.
\end{equation}
In this limiting case we can introduce the continuous limit according to
\begin{equation}
    q = \frac{l \pi}{2 N},\;\;dq = \frac{\pi}{2 N},\;\; q \in \{0,\pi/2\}.
\end{equation}
Then we obtain
\begin{equation}\label{Se_cont}
    S_1^{end} = \frac{1}{\pi}\int_0^\frac{\pi}{2}dq \frac{\cos^2 q}{1+4\kappa \sin^2 q} = \frac{\sqrt{1+4\kappa}-1}{8\kappa}\simeq \frac{1}{4\kappa^{1/2}},\;
\end{equation}
and
\begin{equation}\label{Sc_cont}
    S_1^{mid} = \frac{1}{2\pi}\int_0^\frac{\pi}{2}dq \frac{1}{1+4\kappa \sin^2 q} = \frac{1}{4\sqrt{1+4\kappa}}\simeq \frac{1}{8\kappa^{1/2}}\;.
\end{equation}
Here, we have introduced the activity parameter 
\begin{equation}
    \kappa = \tau\gamma = \frac{\tau}{\tau_m}\;,
\end{equation}
which is the ratio of the persistence time of the direction of the active force and the diffusive monomer relaxation time, $\tau_m$. Since active motion of the monomers should be always dominant we can safely assume $\kappa\gg 1$ which leads to the asymptotic expressions in Eqs.(\ref{Se_cont}) and (\ref{Sc_cont}).  In this limit the exponent $\epsilon=(S_2-S_1)/S_2$ which determines the degree of stationary anti-chemotactic behavior reads
\begin{equation}
    -\epsilon_{end} = \frac{N}{4\sqrt{\kappa}} -1 \simeq -2\epsilon_{mid}\;.
\end{equation}
For chemotactic behavior ($\epsilon <0$) we have to assume $N^2 \gg \kappa$, or $\tau \ll \tau_m N^2 =\tau_R$, which means that the active persistence time should be much smaller than the diffusive relaxation time, or Rouse time, of the whole chain. We note that the relation $|\epsilon_{end}| > |\epsilon_{mid}|$ holds generally using the exact results in Eqs.(\ref{Se_cont}) and (\ref{Sc_cont}) for $\kappa>1$.

Within the same limit can consider the result for the all-active chain, i.e. $\forall \alpha_l =1$, For this case obtain
\begin{equation}
    S_1^{all} =   \sum_{l=1}^{N-1} \frac{1}{1+\tau\gamma\sin^2 \left(\frac{l\pi}{2N}\right)}=N\cdot S^{end}_1\;\;\text{and}\;\;S_2^{all}=1 =N\cdot S_2^{end,mid}\;,
\end{equation}
where we have used the normalization condition of the eigenfunctions in Eq.(\ref{eqn:S1}). Thus, we obtain the large $N$ limit:
\begin{equation}\label{eq:gen_epsilon_relation}
    |\epsilon_{end}| \simeq 2 |\epsilon_{mid}| \simeq 2|\epsilon_{all}|\;\;.
\end{equation}

\section{Mean first passage time (MFPT)}
The coarse grained Fokker-Planck Equation can be rewritten as
\begin{equation}
    \frac{\partial \rho}{\partial t} = - \nabla\cdot \left( \left(1 - \frac{\epsilon}{2} \right) \nabla \mathcal{D} \rho\right)  + \nabla^2 (\mathcal{D} \rho).
\end{equation}
Following the procedure in Sec 5.2 in \cite{gardiner1985handbook}, the equation of the mean first passage time $T(x)$ for a one dimensional problem with absorbing barriers at $a$ and $b$, and $x$ being the initial position of the polymer's center of mass is 
\begin{equation}
   \left( \left(1 - \frac{\epsilon}{2} \right) \partial_x \mathcal{D} \right) \partial_x T(x) + \mathcal{D}\partial^2_x T(x) = -1,
\end{equation}
which can be solved analytically to yield,
\begin{equation}
    T(x) = \frac{\int_a^x \frac{dy}{\psi (y)} \int_x^b \frac{dy'}{\psi (y')} \int_a^{y'}\frac{dz \psi (z)}{\mathcal{D}(z)} - \int_x^b \frac{dy}{\psi (y)} \int_a^x \frac{dy'}{\psi (y')} \int_a^{y'}\frac{dz \psi (z)}{\mathcal{D}(z)}}{\int_a^b \frac{dy}{\psi (y)}},
\end{equation}
where $\psi (x)$ is the integrating factor, given by -
\begin{equation}
    \psi (x) = \exp \left(\int_a^x dx' \frac{ \left(1 - \frac{\epsilon}{2} \right) \partial_x \mathcal{D}(x')}{\mathcal{D}(x')} \right).
\end{equation}
We are interested in calculating the MFPT for a polymer in reaching the location of maximum activity starting from the center of a box with length $L=100$. Since we consider a periodically varying activity field $f_s(x) = 20(1 + \sin(2\pi x/L))$, we apply absorbing boundary conditions at $x=25$ and $125$ i.e $a=25$ and $b=125$. Setting $t_m = T(50)$, we get the required MFPT. 
\end{document}